# Encoding innumerable charge density waves of FeGe into polymorphs of LiFe$_6$Ge$_6$

Yilin Wang[1,2,3]

[1]*School of Emerging Technology, University of Science and Technology of China, Hefei 230026, China*

[2]*Hefei National Laboratory, University of Science and Technology of China, Hefei 230088, China*

[3]*New Cornerstone Science Laboratory, University of Science and Technology of China, Hefei, 230026, China*

**Kagome metals exhibit rich quantum states by the intertwining of lattice, charge, orbital and spin degrees of freedom. Recently, a novel charge density wave (CDW) ground state was discovered in kagome magnet FeGe and was revealed to be driven by lowering magnetic energy via large Ge1-dimerization. Here, based on DFT calculations, we show that such mechanism will yield infinitely many metastable CDWs in FeGe due to different ways to arrange the Ge1-dimerization in enlarged superstructures. Intriguingly, utilizing these metastable CDWs, innumerable polymorphs of kagome magnet LiFe$_6$Ge$_6$ can be stabilized by filling Li atoms in the voids right above/below the dimerized Ge1-sites in the CDW superstructures. Such polymorphs are very stable due to the presence of magnetic-energy-saving mechanism, in sharp contrast to the non-magnetic "166" kagome compounds. In this way, a one-to-one mapping of the metastable CDWs of FeGe to stable polymorphs of LiFe$_6$Ge$_6$ is established. On one hand, the fingerprints of these metastable CDWs, i.e., the induced in-plane atomic distortions and band gaps, are encoded into the corresponding stable polymorphs of LiFe$_6$Ge$_6$, such that further study of their properties becomes possible. On the other hand, such innumerable polymorphs of LiFe$_6$Ge$_6$ offer great degrees of freedom to explore the rich physics of magnetic kagome metals. We thus reveal a novel connection between the unusually abundant CDWs and structural polymorphism in magnetic kagome materials, and establish a new route to obtain structural polymorphism on top of CDW states.**

# 1. Introduction

Kagome metal features geometry frustration [1], flat bands [2-6], magnetism [7-10], non-trivial topology [11-14], van Hove singularities (VHSs) [15,16], strong electron-phonon interactions [17-20]. It thus provides an ideal platform for exploring rich quantum states. In particular, abundant charge density wave (CDW) states have been discovered. For instance, an exotic chiral CDW state was observed [21,22] in $AV_3Sb_5$ (A=K, Rb, Cs) [11,23], and was suggested to be driven by its VHSs [15,16,24-32]. Abundant competing CDW instabilities were observed in $ScV_6Sn_6$ [33-38]. These two systems are non-magnetic with moderate electronic correlations [39]. Recently, a novel CDW was observed in kagome magnet FeGe around 100 K [40-42], which is deep inside its A-type antiferromagnetic (AFM) phase. Remarkably, the ordered magnetic moments are significantly enhanced by the CDW transition [40], indicating possible intertwining of CDW order and magnetism. Intensive investigations have been devoted to uncover its origin [40-50]. An unconventional CDW mechanism driven by saving magnetic energy via the interplay of large Ge1-dimerization along *c*-axis, strong electronic correlations of Fe-*3d* orbitals, and magnetism was revealed for FeGe [42,50]. The key finding is that partial (a quarter of) Ge1-dimerization in a 2 × 2 × 2 superstructure would save magnetic energy, which overcomes the energy cost from this structural distortion and leads to a new ground state, in which a 2 × 2 in-plane CDW ordering is induced to respond to the large partial Ge1-dimerization. Further detailed analysis of the magnetic exchange couplings of FeGe confirms this physical picture [48], and the large partial Ge1-dimerization was recently confirmed by single crystal x-ray diffraction experiments [51,52] with high-quality FeGe sample obtained by annealing treatment [53].

Here, based on density functional theory (DFT) calculations, we further show that such competition between magnetic energy saving and structural energy cost via large dimerization of *all the Ge1-sites* will lead to infinitely many metastable CDW states in FeGe, in addition to the CDW ground state with partial Ge1-dimerization, because there are numerous different ways to arrange the Ge1-dimerization in enlarged superstructures of FeGe. Intriguingly, utilizing these metastable CDW superstructures, we find that it could yield innumerable polymorphs of magnetic kagome metal $LiFe_6Ge_6$ ("166" for short) by filling Li atoms right above/below the dimerized Ge1-sites. Such polymorphs are very stable due to the presence of magnetic-energy-saving mechanism, in sharp contrast to the non-magnetic "166" kagome compounds. Thus, a one-to-one mapping of the innumerable metastable CDWs of FeGe to stable polymorphs of LiFe$_6$Ge$_6$ is established. In this way, the fingerprints of the metastable CDWs of FeGe, i.e., the induced in-plane atomic distortions and band gaps, are encoded into the corresponding stable polymorphs of LiFe$_6$Ge$_6$, such that further study of their properties becomes possible. Moreover, the abundant polymorphs of LiFe$_6$Ge$_6$ enrich the family of "166" kagome magnets and offer great degrees of freedom to explore the rich physics of magnetic kagome metals [5,7-9,12]. We thus reveal a novel connection between the unusually abundant CDWs and structural polymorphism in magnetic kagome materials, and establish a new route to obtain structural polymorphism on top of CDW states. The experimental synthesis of such abundant polymorphs of LiFe$_6$Ge$_6$ and exploration of their potential applications are worthy of further studies.

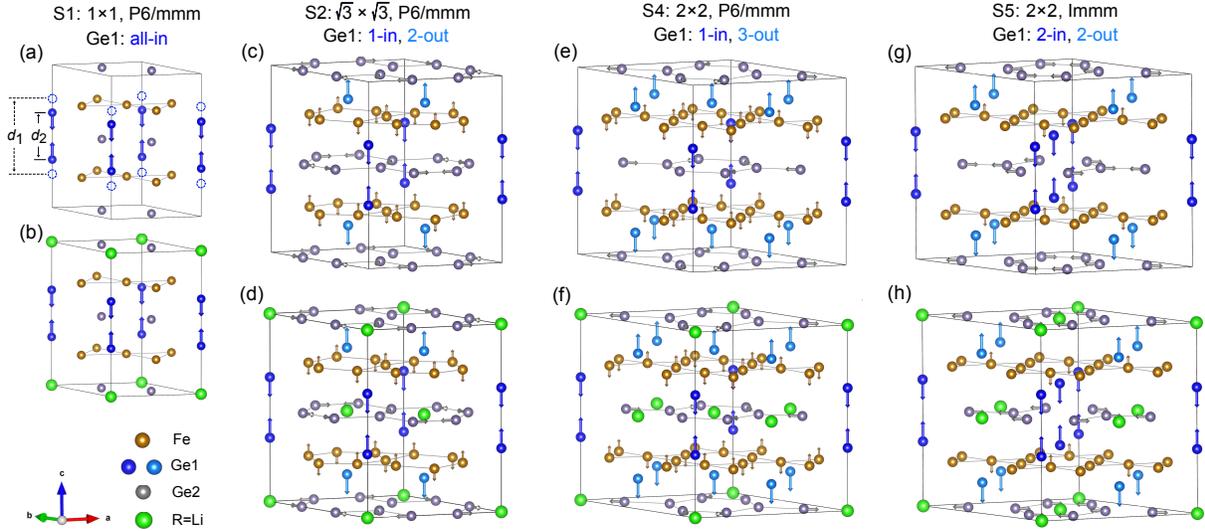

**Figure 1** (Color online) One-to-one mapping of metastable CDWs of FeGe to polymorphs of LiFe$_6$Ge$_6$. The polymorphs of LiFe$_6$Ge$_6$ (second row) are obtained by filling Li atoms (green) into the voids of Ge2-honeycomb layers (grey), right above/below the dimerized Ge1-sites (blue) in the metastable CDW superstructures of FeGe (first row). (a), (b) All the Ge1-sites move inward (relative to the center plane) along the $c$ axis to form dimers. The blue dashed circles indicate the positions of Ge1-sites in the kagome layers before dimerization. The dimerization strength is defined as the Ge1-Ge1 bond length before and after dimerization, $d = d_1 - d_2$. The Li atoms are filled into every two honeycomb layers. (c)-(h) Enlarged superstructures along the crystal $a$ and $b$ axis, with some of Ge1-sites moving inward (dark blue) and the others moving outward (light blue). (c), (d) $\sqrt{3} \times \sqrt{3}$ superstructure with one pair of Ge1-sites moving inward and two pairs of Ge1-sites moving outward ("1-in, 2-out"), where one Li-site at $z = 0$ and two Li-sites at $z = 0.5$. (e), (f) $2 \times 2$ superstructure with "1-in, 3-out". (g), (h) $2 \times 2$ superstructure with "2-in, 2-out". The longer arrows indicate the large dimerization of Ge1-sites ($d \sim 1.3$ Å), and the shorter ones indicate the small distortions of Fe and Ge2 sites (< 0.05 Å). More superstructures are shown in the supplementary Figures S1-S6.

## 2. Methods

The DFT calculations are performed using the VASP package [54,55], with exchange-correlation functional of generalized gradient approximation (GGA) [56]. Although FeGe is a strongly correlated magnet, we have shown that the DFT calculations without Hubbard $U$ have produced the correct ordered magnetic moments of its AFM phase (around 1.5 μ$_B$ /Fe) [50] observed by neutron scattering experiment [40], and predicted the correct CDW ground state. Therefore, DFT calculations are applicable to the AFM phase of FeGe, and it should be also applicable to LiFe$_6$Ge$_6$ since their magnetic and electronic properties are very similar. The experimental lattice parameters, $a$ = 4.985 Å and $c$ = 4.048 Å are used for FeGe [57]. The A-type AFM configuration with in-plane ferromagnetic couplings and inter-layer AFM couplings is considered in the AFM calculations, since our calculations show that it is also the ground magnetic structure for LiFe$_6$Ge$_6$ (see Figure S10 and Table S2 in the Supplementary Information). The energy cutoff of the plane-wave basis is set to be 500 eV. When relaxing the CDW superstructure, the internal atomic positions are relaxed until the force of each atom is smaller than 1 meV/Å. The spin-orbit coupling is not included in all the calculations, since it is very small for Fe and Ge ions and will not change the conclusions of the present work. The phonon spectra are calculated by the finite displacement method with the aid of *phonopy* code [58]. The unfolding of band structures is performed using the *BandUP* code [59]. See more computational details in the Supplementary Information.

## 3. Results

As shown in Figure 1(a), hexagonal FeGe is consisting of kagome layers of Fe-sites (brown) and two non-equivalent Ge-sites: Ge1 in the center of the hexagon of kagome structure (dashed blue circles, before dimerization) and Ge2-honeycomb layers (grey). The Ge1-sites can move away from the kagome layers along the *c*-axis by about 0.65 Å to form Ge1-Ge1 dimers [42,50,51], as indicated by the blue arrows. The dimerization strength is defined as the Ge1-Ge1 bond lengths before and after dimerization, $d = d_1 - d_2$, where $d_1 \sim 4$ Å. It should be noted that only a quarter of Ge1-sites are dimerized in its CDW ground state [42,50,51]. Here, we show that FeGe superstructures in which *all the Ge1-sites are dimerized* can also lead to similar in-plane CDW metastable states. If all the Ge1-sites in one kagome layer move in the same $c$-direction toward the central honeycomb plane, it can only form the $1 \times 1$ structure without any in-plane charge modulations in the kagome and honeycomb layers (Figure **1(a)**). While, it could lead to numerous different superstructures if part of the Ge1-sites move along one $c$-direction and the others move in opposite $c$-direction. For example, as shown in Figure **1(c)**, one pair of Ge1-sites move inward and the other two pairs move outward ("1-in, 2-out" for short) in $\sqrt{3} \times \sqrt{3}$ superstructure. For a $2 \times 2$ superstructure, there are two choices: "1-in, 3-out" (Figure **1(e)**) and "2-in, 2-out" (Figure **1(g)**). Different from the $1 \times 1$ structure, such kinds of Ge1-dimerization will induce in-plane charge modulations, accompanying with small distortions (~0.05 Å) of Fe and Ge2-sites (indicated by the shorter arrows). For example, it induces distortions of Fe-sites along the *c*-axis and a Kekule-type distortion [60] of Ge2-sites in Figure **1(e)**, which are exactly the same distortions as in the CDW ground state of FeGe [42,50,51], because both distorted superstructures hold the same symmetry (space group P6/mmm). Therefore, infinite number of metastable CDW states could be obtained in this way, thanks to the magnetic-energy-saving mechanism by Ge1-dimerization [50]. To demonstrate this, we consider 30 different FeGe superstructures of such kinds (labeled as S1-S30, shown in Figure **1** and the supplementary Figures S1-S6). It includes superstructures of $1 \times 1$ (1), $\sqrt{3} \times \sqrt{3}$ (1), $1 \times 3$ (1), $2 \times 2$ (2), $1 \times 4$ (2), $1 \times 5$ (3), $1 \times 6$ (5), $2 \times 3$ (5), $1 \times 8$ (4) and $3 \times 3$ (6), where the number of polymorphs are indicated in the parentheses.

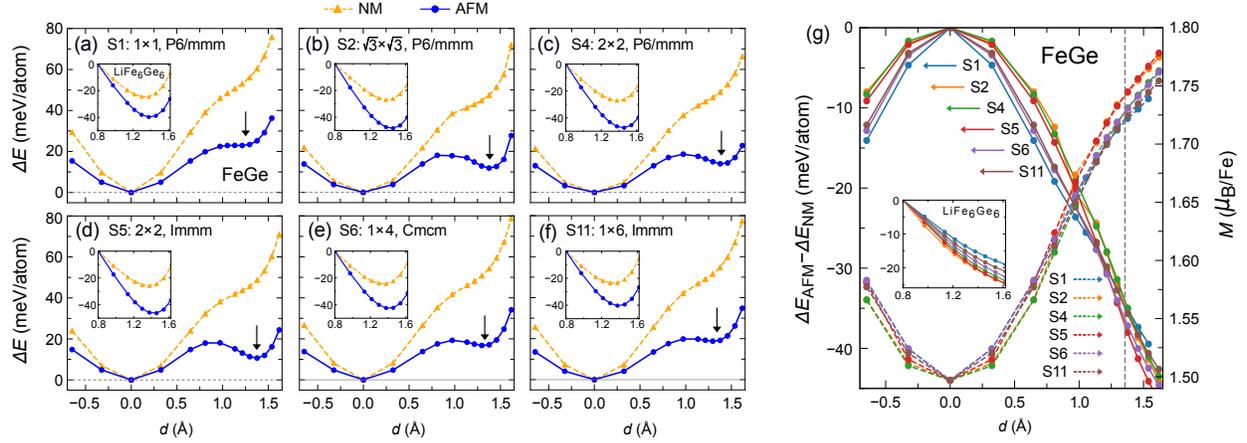

**Figure 2** (Color online) Innumerable metastable CDWs driven by magnetic-energy-saving in FeGe. (a)-(f) The DFT calculated total energies, $\Delta E = E(d) - E(d=0)$, of the AFM (solid blue) and non-magnetic states (dashed orange) as functions of Ge1-dimerization strength $d$ for 6 typical metastable CDW superstructures of FeGe. The insets are the total energies, $\Delta E = E(d) - E(d=0.8)$, as functions of $d$ for the corresponding polymorphs of LiFe$_6$Ge$_6$. The black arrows indicate the local energy minimums around $d \sim 1.3$ Å in FeGe, which become global energy minimums in polymorphs of LiFe$_6$Ge$_6$. (g) Left-$y$ axis: the energy differences between the AFM and non-magnetic states that measures the magnetic exchange energies (solid), as functions of $d$ for metastable CDW superstructures of FeGe, and the inset is for the corresponding polymorphs of LiFe$_6$Ge$_6$. Right-$y$ axis: the ordered magnetic moments of Fe (dashed) as functions of $d$ for metastable CDWs of FeGe.

Figures **2(a)-(f)** show the DFT calculated total energies, $\Delta E = E(d) - E(d=0)$, as functions of the averaged Ge1-dimerization strength $d$ for 6 typical FeGe superstructures. Results for more superstructures are shown in the supplementary Figure S7. The solid blue and dashed orange curves are for the AFM and non-magnetic (NM) states, respectively. $\Delta E$ keeps increasing with $d$ in the NM state, indicating that it has to pay for energy for the structural distortions of Ge1-dimerization. The energy differences between the AFM and NM states, $\Delta E_{AFM} - \Delta E_{NM}$, that measures the magnetic exchange energies, are negative and decrease with increasing $d$ (Figure **2(g)**, left *y*-axis), and the ordered magnetic moments (Figure **2(g)**, right *y*-axis) increase with $d$. These indicate the presence of magnetic-energy-saving by dimerizing all the Ge1-sites in these superstructures, similar to that in the CDW ground state of FeGe with partial Ge1-dimerization [50]. The competition between magnetic energy saving and structural energy cost result in local energy minimums around $d \sim 1.3$ Å (indicated by the black arrows), confirming the formation of metastable CDW superstructures.

Voids are left right above/below the dimerized Ge1-sites in these metastable CDW superstructures, such that another atom $R$ can be filled into the center of the hexagon of Ge2-honeycomb layers (right above/below Ge1-sites), as illustrated by the green color in Figures **1(b), (d), (f), (h)**. Here, we choose $R$=Li since it has the smallest atomic radius. This results in numerous polymorphs of kagome magnet LiFe$_6$Ge$_6$ (see Figure **1** and the supplementary Figures. S1-S6). It is noted that most of the "166" compounds crystallize into the $1 \times 1$ structure (Figure **1(b)**), where the $R$ atoms are present in every two honeycomb layers. While, for the polymorphs of LiFe$_6$Ge$_6$, the structures are enlarged along the *a* and *b* axis with a certain number of Li atoms present in each honeycomb layer. The Li atoms act as a skeleton to further stabilize the dimerized metastable CDW superstructures via their crystalline electric field. As shown in the insets of Figures **2(a)-(f)**, the local energy minimums around $d \sim 1.3$ Å in FeGe become global energy minimums in LiFe$_6$Ge$_6$. We note that the magnetic energy saving from Ge1-dimerization is still effective in the polymorphs of LiFe$_6$Ge$_6$ and plays important roles in stabilizing the structures, in addition to the crystalline electric field from Li-sites. This is shown in the inset of Figure **2(g)**, where $\Delta E_{AFM} - \Delta E_{NM}$ decrease with increasing $d$, indicating the presence of magnetic-energy-saving. Therefore, the polymorphs of magnetic LiFe$_6$Ge$_6$ should be very stable. This is in sharp contrast to the non-magnetic "166" kagome compounds, where such kind of magnetic-energy-saving mechanism is absent, as shown by the non-magnetic calculations in Figures **2(a)-(f)**. Consequently, the crystalline electric field from $R$ sites is the only factor to stabilize the dimerized structures in non-magnetic "166" kagome compounds, such that further movement of the dimerized sites along the $c$-axis could be possible. For example, in non-magnetic ScV$_6$Sn$_6$ that crystallizes into the $1 \times 1$ type structure [see Figure **1(b)**] at high temperature, further distortions of the dimerized Sn sites along the $c$-axis induce a $\sqrt{3} \times \sqrt{3} \times 3$ CDW transition at 92 K [33-38].

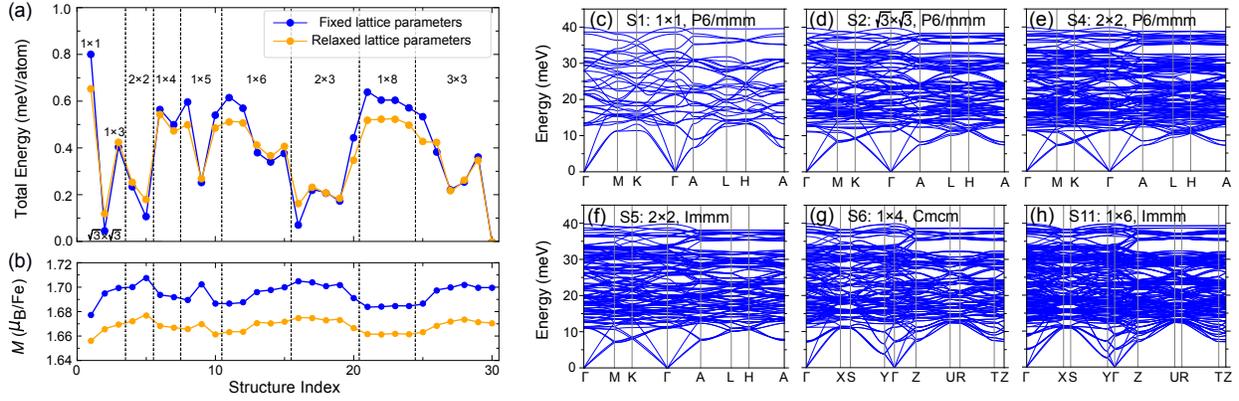

**Figure 3** (Color online) Innumerable stable polymorphs of LiFe$_6$Ge$_6$. The DFT calculated total energy with respect to the lowest one (S30) in (a) and ordered magnetic moments in (b) of the 30 typical polymorphs of LiFe$_6$Ge$_6$ with experimentally fixed (blue) and DFT-relaxed (orange) lattice parameters. (c)-(h) The DFT calculated phonon spectra of 6 typical polymorphs of LiFe$_6$Ge$_6$. A hexagonal Brillouin zone (BZ) is used in (c)-(f), and an orthorhombic BZ is used in (g) and (h).

The DFT calculated total energies of the 30 polymorphs of LiFe$_6$Ge$_6$ (S1-S30) are shown in Figure **3(a)**. Blue and orange curves show the results for the fixed and relaxed lattice parameters, respectively. The fixed lattice parameters are extracted from a previous experiment [61], where the $\sqrt{3} \times \sqrt{3}$ superstructure (S2) of LiFe$_6$Ge$_6$ has already been successfully synthesized. A 3 × 3 superstructure (S30) is predicted to be lowest in energy among the 30 polymorphs and is very close to that of the $\sqrt{3} \times \sqrt{3}$ superstructure (S2) in our calculation. The energy differences of all the 30 polymorphs are very small (within 1 meV/atom), suggesting that these polymorphs (at least some of them) could be synthesized in suitable experimental conditions. The calculated ordered magnetic moments are about 1.7 μ$_B$/Fe [Figure **3(b)**], which are close to the values of the metastable CDW superstructures of FeGe (indicated by the dashed vertical line in Figure **2(g)**). The calculated phonon spectra for several typical polymorphs LiFe$_6$Ge$_6$ are shown in Figures **3(c)-(h)** and the supplementary Figure S8. No obvious soft phonon modes can be identified, which further confirms the stability of these polymorphs due to magnetic-energy-saving. We also performed *ab initio* molecular dynamics (AIMD) simulations at $T$ = 300 K for three representative polymorphs of LiFe$_6$Ge$_6$, including 1×1 (S1), $\sqrt{3} \times \sqrt{3}$ (S2), 1×4 (S6, Cmcm). The total energies as functions of simulation times are shown in Figure S11. The results show that the polymorphs are largely maintained as its initial structures during the AIMD simulations up to 5 ps, indicating they are also thermally stable and will not decompose on its own.

Filling Li into the voids of the metastable CDW superstructure of FeGe does not change its space group and symmetry. Therefore, the diverse atomic distortion patterns in the kagome and honeycomb layers associated with the abundant metastable CDWs of FeGe, have been retained and encoded into these stable polymorphs of LiFe$_6$Ge$_6$. Further study of their properties such as their electronic structures becomes possible. Figure **4** compares the DFT calculated band structures between the polymorphs of LiFe$_6$Ge$_6$ and the corresponding metastable CDWs of FeGe for $\sqrt{3} \times \sqrt{3}$ (P6/mmm), 2 × 2 (P6/mmm) and 2 × 2 (Immm) superstructures, respectively. The band structures of LiFe$_6$Ge$_6$ are very similar to those of FeGe, and can be obtained by an approximate rigid shift of the band structures of FeGe downward by 0.12 eV. This is because, apart from providing one extra valence electron into the system, the Li ion will barely affect the dispersion of the electronic bands. Band gaps induced by metastable CDWs of FeGe can be found around the Fermi level, as indicated by the red rectangles in Figures **4(b)-(d)**. Such fingerprints of CDWs are also retained in the corresponding polymorphs of LiFe$_6$Ge$_6$ around 0.12 eV below the Fermi level, as indicated by the red rectangles in Figures **4(f)-(h)**, which can be observed by angle-resolved photoemission experiment (ARPES).

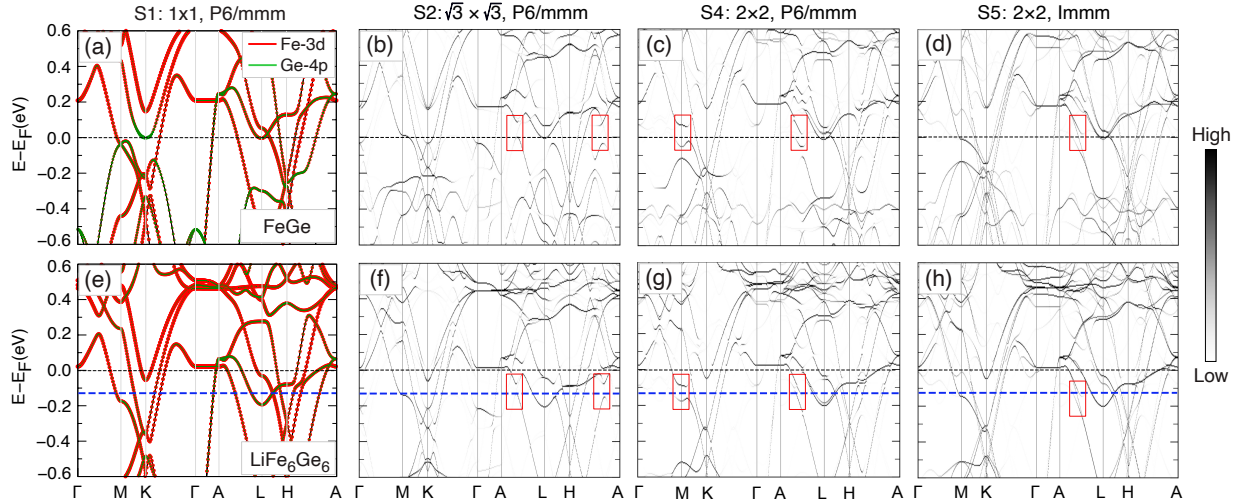

**Figure 4** (Color online) Encoding the fingerprints of metastable CDWs of FeGe into stable polymorphs of $LiFe_6Ge_6$. (a)-(d) Band structures for metastable CDWs of FeGe, (e)-(h) and those for polymorphs of $LiFe_6Ge_6$. The band structures are calculated in the A-type AFM magnetic configuration. The band structures of $\sqrt{3} \times \sqrt{3}$ and $2 \times 2$ superstructures are unfolded into the BZ of the $1 \times 1$ structure. The band structures of $LiFe_6Ge_6$ can be obtained by an approximate rigid shift of the band structures of FeGe downward by 0.12 eV, as indicated by the blue dashed lines. The red rectangles indicate the band gaps around the Fermi level induced by metastable CDWs of FeGe, which are retained in the corresponding polymorphs of $LiFe_6Ge_6$ around -0.12 eV.

## 4. Discussion and summary

To summarize, we have demonstrated that infinitely many metastable CDWs can be driven by magnetic-energy-saving in enlarged superstructures of FeGe by dimerizing all the Ge1-sites in different ways. Utilizing such abundant metastable CDW superstructures of FeGe, we show that innumerable polymorphs of kagome magnet $LiFe_6Ge_6$ could be stabilized by the same magnetic-energy-saving mechanism. We thus reveal a novel connection between CDWs and structural polymorphism in this magnetic kagome system. In this way, the fingerprints of the abundant metastable CDWs of FeGe, i.e., the induced in-plane atomic distortions and band gaps, are encoded into the corresponding stable polymorphs of $LiFe_6Ge_6$. In this sense, the polymorphs of $LiFe_6Ge_6$ serve as "mimic" of the metastable CDWs of FeGe. We thus establish a new route to obtain abundant structural polymorphism on top of CDW states.

Considering that the energy differences of the polymorphs of LiFe$_6$Ge$_6$ are very small, we expect that at least part of them can be synthesized in suitable experimental conditions. Many polymorphs may even coexist in one synthesized sample. We note that it may be challenging to synthesize those polymorphs with relatively larger total energies. Since the crystal structures of LiFe$_6$Ge$_6$ polymorphs are determined by the strengths and directions of Ge1-dimerization, so it is essential to tune the Ge1-dimerization when synthesizing the polymorphs experimentally. Recent experiment [53] found that annealing treatment at different temperatures can effectively tune the Ge1-dimerization in FeGe, such that both the ground and some metastable CDW states could be realized. Therefore, annealing treatment at different temperatures is a feasible and promising way to stabilize these LiFe$_6$Ge$_6$ polymorphs with relatively larger energies, at least, in some regions/domains of the samples. Since the polymorphs are determined by the positions of Li-sites and the atomic radius of Li is small, the transition between different polymorphs may be also achieved. Furthermore, how to utilize such polymorphic transition in LiFe$_6$Ge$_6$ to achieve some functionality deserves further studies.

Our theoretical analysis also suggests that such kinds of stable polymorphs should universally exist in other $R$Fe$_6$X$_6$ compounds ($R$=Mg, Sc, Y, Ti, Zr, Hf, Nb and rare-earth elements, $X$=Ge, Sn), due to the same magnetic-energy-saving mechanism. The total energies of some of the polymorphs of $R$Fe$_6$Ge$_6$ with $R$=Mg, Sc, Lu are also very close. This further enrich the family of magnetic "166" kagome compounds and offers great degrees of freedom to explore the rich physics of magnetic kagome metals.

**Conflict of interest**

The authors declare that they have no conflict of interest.

**Acknowledgments**

We thank Yajun Yan and Aifeng Wang for very helpful discussion. This project was supported by the National Natural Science Foundation of China (No. 12174365) and the New Cornerstone

**Data availability**

All the data supporting the findings of this study are provided within the article and its Supplementary Information files. All the raw data generated in this study are available from the corresponding author upon reasonable request.